\documentclass[a4paper,11pt]{article}
\usepackage{pos}
\usepackage{comment}

\title{Microquasar remnants as reservoirs of PeV cosmic rays}

\author*[a,b]{Leandro Abaroa}
\author[a,b]{Gustavo E. Romero}
\author[c]{Valentí Bosch-Ramon}

\affiliation[a]{Instituto Argentino de Radioastronom\'ia (CCT La Plata, CONICET; CICPBA; UNLP),\\ C.C.5, (1894) Villa Elisa, Argentina}

\affiliation[b]{Facultad de Ciencias Astron\'omicas y Geof\'{\i}sicas, Universidad Nacional de La Plata,\\ B1900FWA La Plata, Argentina}

\affiliation[c]{Departament de F\'isica Qu\`antica i Astrof\'isica, Institut de Ci\`encies del Cosmos (ICC), Universitat de Barcelona (IEEC-UB),\\ Mart\'i i Franqu\`es 1, E08028 Barcelona, Spain}

\emailAdd{leandroabaroa@gmail.com}
\emailAdd{gustavo.esteban.romero@gmail.com}
\emailAdd{vbosch@fqa.ub.edu}

\abstract{
The Large High Altitude Air Shower Observatory (LHAASO) has revealed a population of Galactic $\gamma$–ray sources radiating beyond $100$~TeV, but the nature of several of them is still uncertain. In this contribution, we explore the idea that some of these ultrahigh–energy emitters are not powered by currently active accelerators, but by the fossil remains of microquasars (MQs). We consider systems in which mass transfer onto the stellar–mass black hole has already stopped, so that the central engine and its jets are permanently quenched. During the active phase, powerful transrelativistic jets inflate a hot cocoon whose interior is filled with cosmic rays (CRs) accelerated at the jet termination shocks. Once the jets switch off, the cocoon enters a long "afterlife" stage in which it behaves as a large reservoir of PeV CRs. If the remnant lies in or near a star–forming region, these relic CRs can still interact with dense clumps and molecular clouds, inside the cocoon or in the surrounding interstellar medium, leading to delayed $\gamma$–ray emission via inelastic $pp$ collisions and the subsequent decay of neutral pions. We present a time–dependent model for the jet–cocoon system, follow the evolution of the CR population during and after the MQ phase, and discuss the conditions under which the resulting microquasar remnants can account for some of the unidentified LHAASO sources.}

\FullConference{
High Energy Phenomena in Relativistic Outflows IX (HEPRO-IX)\\
4-8 August 2025\\
Rio de Janeiro, Brazil\\}

\begin{document}
\maketitle

\section{Introduction}

The discovery of Galactic $\gamma$–ray sources emitting up to, and in some cases beyond, the PeV range has revitalized the long–standing quest for the origin of the highest–energy cosmic rays (CRs) in the Milky Way. The recent LHAASO catalog includes a number of ultrahigh–energy (UHE) emitters whose spectra extend above $100$~TeV and in several cases approach or exceed $1$~PeV \citep{LHAASO-Catalog2024ApJS}. The identification of the underlying accelerators---the Galactic PeVatrons---remains challenging, as in many cases no obvious counterpart is seen at lower energies (see, e.g., \cite{2026MNRAS.tmp..261M}).
Energetic arguments indicate that transrelativistic outflows with kinetic power $\gtrsim 10^{39}\,{\rm erg\,s^{-1}}$ are particularly promising candidates \citep{Wang2025ApJ}. A growing body of work shows that microquasars (MQs) and X–ray binaries accreting at super–Eddington rates are able to drive such powerful jets and winds, and can accelerate particles well into the PeV domain \citep{Heinz&Sunyaev_2002,2005A&A...429..267B,2008A&A...485..623R,Abaroa_etal2024(S26),Peretti2025,Zhang2025arXiv250620193Z}. These systems naturally produce large–scale bubbles and cocoons analogous to those observed in radio galaxies, but on galactic scales and with much shorter evolutionary times.

Most discussions so far have focused on MQs that are currently active. However, once mass transfer from the companion star ceases, the central engine can be shut down irreversibly, leaving behind an apparently dormant binary and a fossil cocoon filled with CRs. In this work, we argue that such microquasar remnants (MQRs) can remain efficient CR reservoirs for a long time after the jets have turned off, and that their delayed illumination of nearby dense material may power some of the unidentified LHAASO sources. In this work we present additional results complementing our recent article, in which the MQR framework was introduced \cite{Abaroa_etal_2026_MQR}.

The paper is organized as follows. Sect.~\ref{sect: MQR} summarizes the evolutionary path from an active MQ to an MQR and introduces our jet–cocoon model. Sect.~\ref{sect: relativistic particles} describes the production and transport of CRs in the system, and the resulting proton distributions. Finally, Sect.~\ref{sect: conclusion}  presents the expected emission of an irradiated cloud, discusses the implications of our results, and concludes.

\section{From active microquasars to fossil cocoons}\label{sect: MQR}

\subsection{Pathways to microquasar remnants}

We consider a stellar–mass black hole (BH) of mass $M_{\rm BH}=10\,M_{\odot}$ in a binary system that undergoes a prolonged phase of super–Eddington accretion. During this stage, matter transferred from the companion star feeds a thick accretion flow capable of launching powerful jets and winds. The central engine operates for a time $t_0 = 5\times10^{4}\,{\rm yr}$, representative of the lifetime of a high–mass X–ray binary undergoing sustained mass transfer.

The transition to an MQR occurs once mass transfer is permanently quenched: once the disk is drained, the jets disappear, and the system enters a fossil phase. Several channels can lead to this outcome (see \cite{Ohsuga2005,Belczynski_etal_2016(Nature),Abaroa_etal_2023,Willcox_etal_2023,Abaroa&Romero_RevMex_2024}, and \cite{Abaroa_etal_2026_MQR}).

What survives is a non-interacting binary lying at the center of a large, overpressured cavity inflated by the jets during the active phase. This structure, which we refer to as the MQR cocoon, resembles on smaller scales the lobes of Fanaroff–Riley II radio galaxies \citep{Begelman&Cioffi_1989}, and is indeed observed in some MQs \citep{2009A&A...497..325B,2010Natur.466..209P,Marti_etal_2017NatCo,Abaroa_etal2024(S26)}.

\subsection{Jet launching and termination}

We assume that the kinetic luminosity per jet is $L_{\rm j}=5\times10^{39}\,{\rm erg\,s^{-1}}$, and adopt a mildly relativistic bulk Lorentz factor, $\gamma_{\rm j}=3$, corresponding to a speed $v_{\rm j}\simeq 0.94\,c$ \citep{Heinz&Sunyaev_2002}. The jets propagate through an approximately homogeneous ISM with density $\rho_{\rm ISM}=1.7\times10^{-25}\,{\rm g\,cm^{-3}}$. The advance speed of the jet head is governed by the balance between jet momentum and external ram pressure; following \cite{1997MNRAS.286..215K}, the position of the jet head at time $t<t_0$ can be written as $l_{\rm j}(t) \simeq ( L_{\rm j/}\rho_{\rm ISM} )^{1/5}\, t^{3/5}$. At $t=t_0$, the jet has reached $l_{\rm j}\approx 1.6\times10^{20}\,{\rm cm}\approx 54\,{\rm pc}$. The longitudinal advance speed is $v_{\rm l}={\rm d}l_{\rm j}/{\rm d}t = (3/5)l_{\rm j}/t$, giving $v_{\rm l}\approx 6\times10^{7}\,{\rm cm\,s^{-1}}$ at $t_0$.

At the termination region, a strong reverse shock (RS) decelerates the jet material. This RS is an efficient particle accelerator: it converts a fraction of the jet power into non–thermal particles, which are then advected into the cocoon by the backflow. Observations of non–thermal lobes in MQs and of large–scale jet hotspots in radio galaxies support this picture \citep{2010Natur.466..209P}.

As soon as accretion ceases, the jets vanish and the RS disappears. The previously accelerated particles, however, remain confined in the cocoon for a much longer time.

\subsection{Growth and fossil evolution of the cocoon}\label{sect: cocoon}

The cocoon is inflated by the shocked jet plasma and remains overpressured with respect to the ambient ISM. For an approximately uniform external medium, the cocoon can be represented as an expanding ellipsoid whose major axis follows the jet length during the active phase, $l_{\rm c}(t<t_0)\approx l_{\rm j}(t)$. The transverse size is set by the lateral expansion of the shocked plasma and we adopt a semi–minor axis $w_{\rm c}\simeq l_{\rm c}/3$, consistent with analytical and observational studies of jet–driven lobes \citep{1997MNRAS.286..215K}. The resulting cocoon volume is $V_{\rm c}(t) \simeq 4\pi l_{\rm c}^{3}/27.$

When the jets turn off at $t_0$, the cocoon still contains the energy injected during the active phase, $E_{\rm inj}\simeq 2L_{\rm j}t_0$ (two jets). The subsequent evolution resembles that of a wind–blown bubble that has experienced an impulsive injection of energy. In this stage, the longitudinal size grows as \citep{castor_etal_1977} $l_{\rm c}(t>t_0) \simeq (L_{\rm j} t_0/\rho_{\rm ISM})^{1/5}\, t^{2/5}$, with an expansion speed $v_{\rm c} = {\rm d}l_{\rm c}/{\rm d}t$. The corresponding semi–minor axis and volume follow from the assumed aspect ratio. Even for conservative assumptions, the cocoon reaches sizes of order $100$~pc on timescales of $\sim t_0$ and continues to expand thereafter, although at slower speeds. 


\section{Production and transport of cosmic rays}\label{sect: relativistic particles}

\subsection{Injection at the reverse shock}

We assume that a fraction $q_{\rm rel}=0.1$ of the jet power is converted into CRs at the RS, so that the non–thermal luminosity associated with both jets is $L_{\rm rel}=2q_{\rm rel}L_{\rm j}$. This power is shared between protons and electrons, $L_{\rm rel}=L_{\rm p}+L_{\rm e}$. Motivated by observations and theory of diffusive shock acceleration in collisionless shocks, we adopt a hadron–dominated scenario, where $L_{\rm p}=100\,L_{\rm e}$: about $99$~per cent of the CR power goes into protons and only $1$~per cent into electrons \citep[e.g.,][]{2008A&A...485..623R,caprioli&Spitkovsky2014_PIC}.

Particles are accelerated via first–order Fermi processes at the RS. We estimate the downstream magnetic field by assuming that a fraction $\sim 0.1$ of the thermal pressure is carried by magnetic fields, which gives values of order $B\sim 10\,\mu{\rm G}$. This is only a few times larger than the compressed ISM field and remains well below equipartition with the jet power, so it does not require extreme magnetization. Under these conditions, the maximum proton energy at $t_0$ reaches $E_{\rm max,p}\approx 10^{16}\,{\rm eV}$.

Upstream of the shock, the field is weaker and diffusion is faster, so confinement is mainly controlled by the downstream region. For the purpose of estimating $E_{\rm max,p}$, we therefore use the downstream $B$. Additional acceleration processes, such as second–order Fermi mechanisms inside the cocoon or reacceleration in the shell, are neglected here so as to isolate the role of the RS as the primary accelerator.

\subsection{Transport in the jet termination region and the cocoon}

The distribution of relativistic protons in the RS region, $n_{\rm RS}(E,t)$, is obtained by solving the time–dependent transport equation
\begin{equation}\label{eq: transport_mq}
   \frac{\partial n_{\rm RS}(E,t)}{\partial t}
 + \frac{\partial}{\partial E}
   \left[\dot{E}(E,t)\,n_{\rm RS}(E,t)\right]
 + \frac{n_{\rm RS}(E,t)}{t_{\rm esc,RS}(E,t)}
 = Q_{\rm RS}(E,t),
\end{equation}
where $\dot{E}={\rm d}E/{\rm d}t$ accounts for radiative and adiabatic losses, $t_{\rm esc,RS}$ is the escape timescale from the acceleration region, and $Q_{\rm RS}(E,t)$ is the injection function. We adopt
\begin{equation}
Q_{\rm RS}(E,t) = Q_0(t)\,E^{-p}\exp[-E/E_{\rm max}(t)],
\end{equation}
with index $p=2$, characteristic of standard diffusive shock acceleration. The normalization $Q_0(t)$ is set by the available non–thermal power.

Protons leave the acceleration region predominantly by advection into the backflow with a timescale $t_{\rm esc,RS}\equiv t_{\rm adv}\simeq 4\Delta x/v_{\rm j}$, where $\Delta x$ is the size of the RS region. Once they enter the cocoon, their subsequent evolution is governed by energy losses and spatial diffusion. 
The particle diffusion coefficient can be written as \citep{Ptuskin2012,Peretti2025}

\begin{equation}
D(E) \approx \frac{1}{3}\,c\,L_{\mathrm{c}}
\left( \frac{r_{\mathrm{L}}(E)}{L_{\mathrm{c}}} \right)^{\!\delta}
\left( \frac{B}{\delta B} \right)^{\!2}.
\label{eq:diffusion_general}
\end{equation}
Here, $r_{\mathrm{L}}(E) = E/eB$ is the Larmor radius of a relativistic particle with energy $E$, $B$ is the mean magnetic field, $\delta B$ its turbulent component, and $L_{\mathrm{c}}$ the coherence (or injection) scale of turbulence, where most of the magnetic energy is contained. In the Bohm limit $(\delta = 1)$ the diffusion coefficient reaches its minimum value. We assume a turbulent cocoon under a Kraichnan diffusion regime $(\delta=0.5)$ and adopt $L_{\mathrm{c}} \sim 10^{-2}\,l_{\rm c}$ at each time $t$, corresponding to the expected correlation length of turbulence within the cocoon.

The cocoon thus behaves as a large, slowly leaking CR reservoir. A second transport equation, similar in structure to Eq.~\eqref{eq: transport_mq} but including spatial diffusion and adiabatic losses associated with the expansion of the cocoon, is solved numerically to obtain the proton distribution $n_{\rm c}(E,t)$ throughout the MQ and MQR phases. From that solution, we can compute the total number of protons in the RS region and in the cocoon as functions of time (see \cite{Abaroa_etal_2026_MQR}).

\begin{figure}
    \centering
    \includegraphics[width=7.5cm]{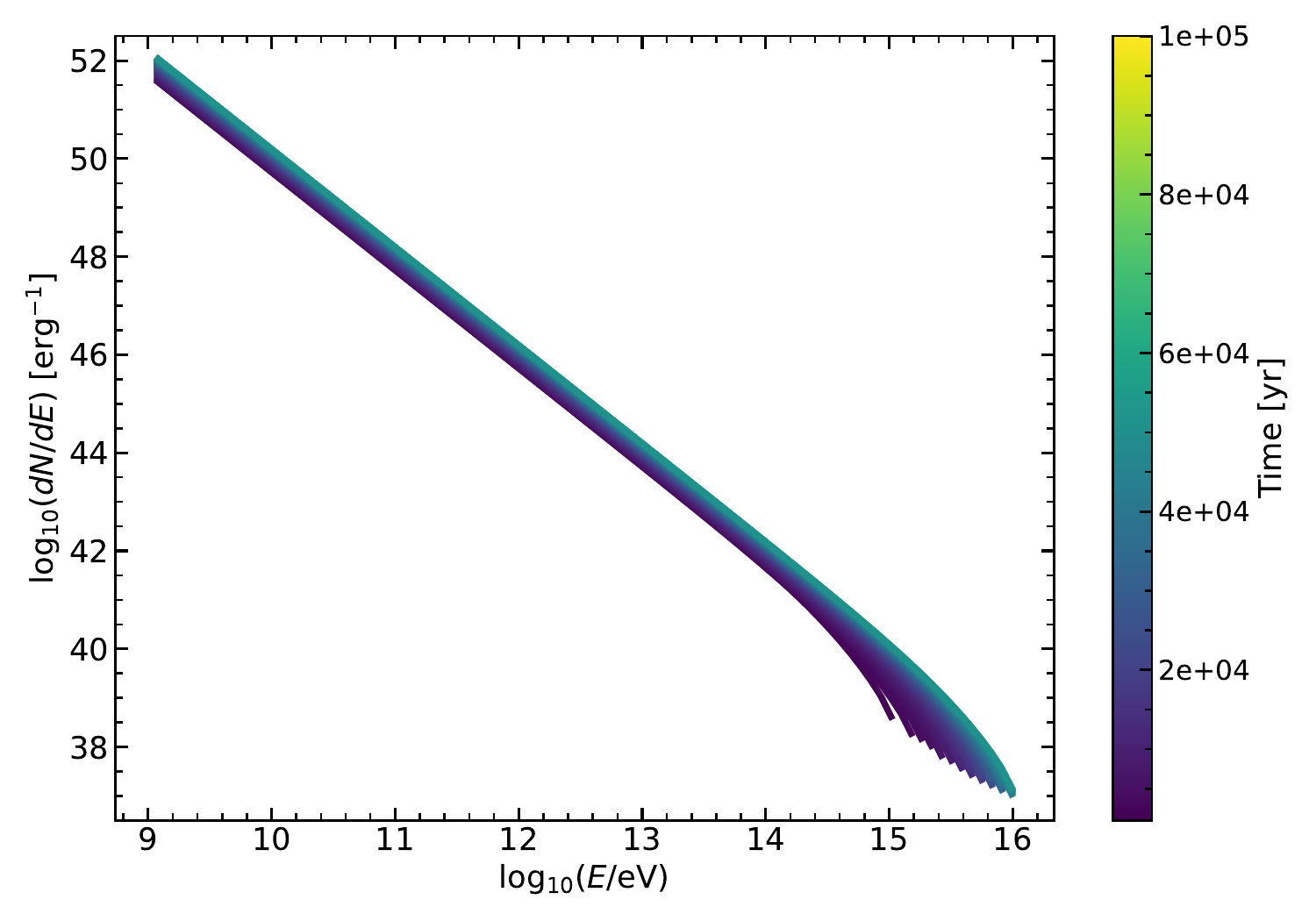}
    \includegraphics[width=7.5cm]{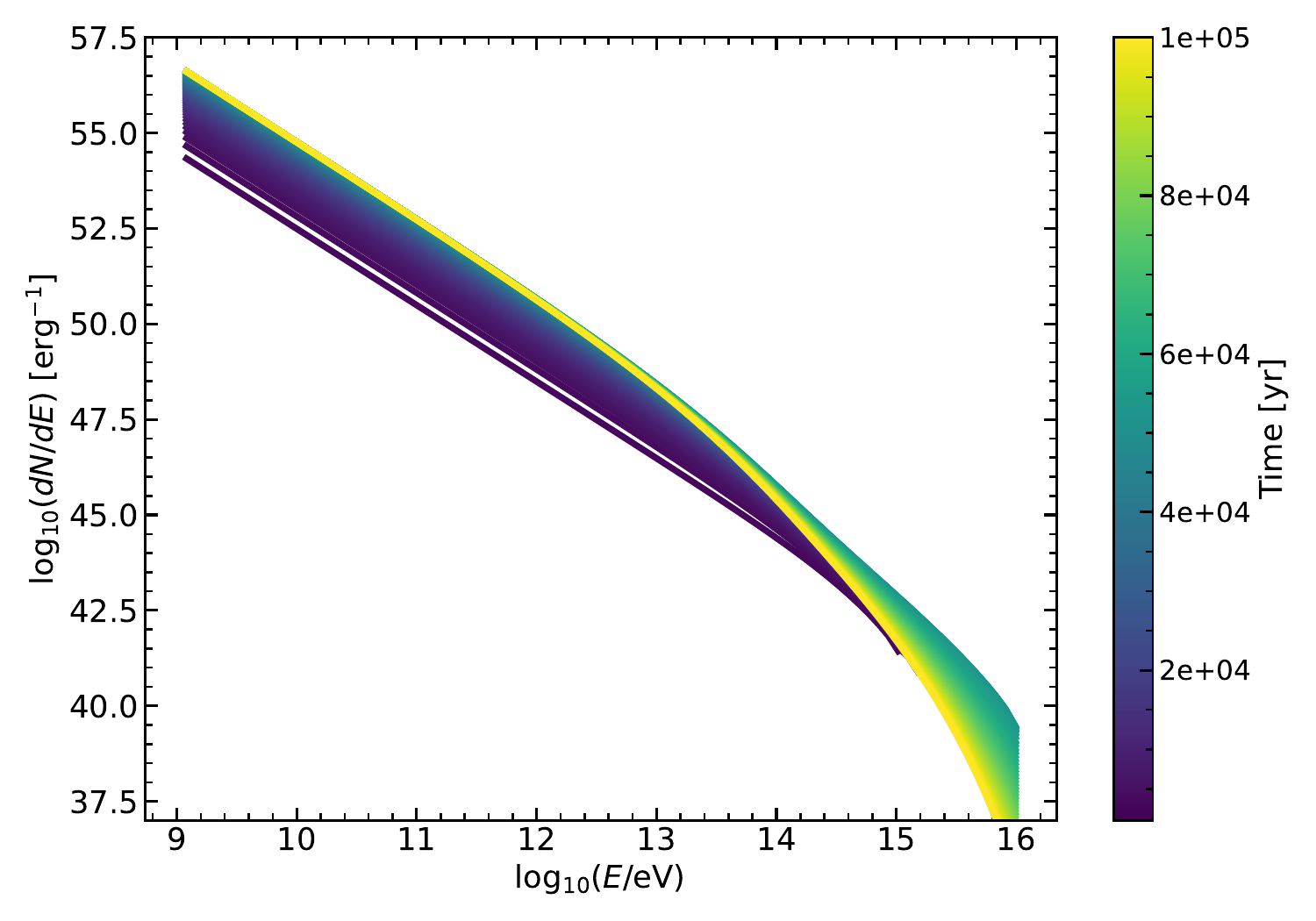}
    \caption{\footnotesize Total number of protons as a function of time in the jet termination region (\textit{left}) and in the cocoon (\textit{right}). The color scale encodes the elapsed time since the onset of the MQ phase. Particles quickly accumulate in the cocoon, which retains a substantial CR population long after the jets have switched off.}
    \label{fig: distribution}
\end{figure}

Figure~\ref{fig: distribution} shows the evolution of the total number of protons in the jet termination region (left panel) and in the cocoon (right panel). During the active phase, particles are rapidly transferred from the RS to the cocoon, so that the latter soon contains the bulk of the CR population. After the jets switch off at $t_0$, the proton content of the cocoon decreases only gradually, reflecting the long diffusion times in the turbulent interior. In spite of this slow leakage, particles with energies $\gtrsim 1\,{\rm PeV}$ persist in the cocoon hundreds of thousands of years after the central engine has died.

\section{Discussion and conclusions}\label{sect: conclusion}

Once the MQ becomes an MQR, no fresh particles are injected, but the cocoon still contains a substantial reservoir of CRs. These particles can interact with dense gas in two distinct environments: (i) clumps and fragments of molecular clouds engulfed by the expanding cocoon, and (ii) clouds situated outside the cocoon, which are illuminated by CRs that have escaped into the ISM.

The timescale for CRs to reach a cloud at distance $R$ from the MQR center is approximately $\tau_{\rm diff}\sim R^{2}/D(E)$, so that the higher–energy particles arrive first. As a result, the $\gamma$–ray spectrum from a given cloud is both space– and time–dependent, and the highest energy photons are emitted preferentially at early times.

Because the cocoon can encompass several clumps and because the escaping CRs can illuminate multiple external clouds, a single MQR has the potential to generate a small cluster of $\gamma$–ray sources. These sources would display different flux levels and spectral shapes, depending on their distance from the MQR and on the local gas density. Energy–dependent diffusion naturally leads to spectral steepening with increasing distance from the CR reservoir. This offers a way to constrain the diffusion coefficient in the vicinity of the remnant.

The picture emerging from our modeling is that of an extended, low–surface–brightness  object, acting as a long–lived PeVatron long after the demise of its central engine. The cocoon of an MQR can reach sizes $\sim 100$~pc and remain overpressured and approximately adiabatic over timescales $t\gtrsim 10^{5}\,{\rm yr}$, and protons with energies above $\sim 1$PeV remain confined for a substantial fraction of this time.

\begin{figure}
    \centering
    \includegraphics[width=0.7\linewidth]{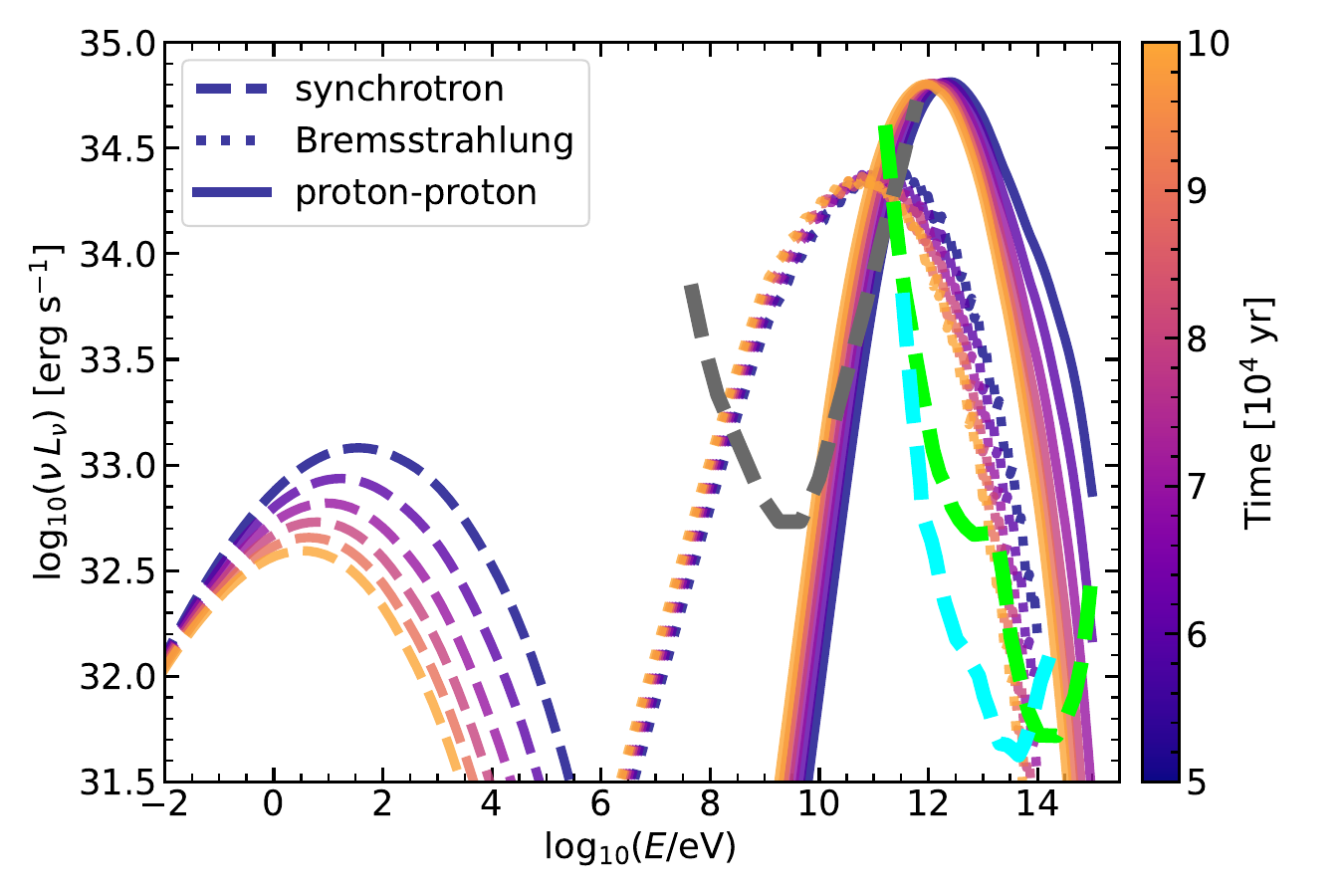}
    \caption{\footnotesize Spectral energy distributions of the illuminated cloud at 50 pc from the MQR, at different times. Leptonic emission is produced by secondary pairs created in the cloud via $pp$ interactions. Thick dashed lines show the sensitivity curves of \textit{Fermi} (10 yr, gray), SWGO (5 yr, light blue), and LHAASO (1 yr, green) for a Galactic MQR at 3 kpc.}
    \label{fig:sed}
\end{figure}

From an observational perspective, the clearest tracers of MQRs are likely to be $\gamma$–ray sources associated with irradiated clouds. Clouds located inside the cocoon or in its immediate surroundings would appear as very-high energy (VHE) or UHE $\gamma$–ray emitters, while the central accelerator may remain largely undetected at other wavelengths. As an illustrative example, we show in Fig. \ref{fig:sed} the time-dependent spectral energy distribution of a cloud illuminated by an MQR. The cloud is located at 50 pc from the MQR, has a radius of 10 pc, a density of $10^3\,{\rm cm^{-3}}$, and a magnetic field of $\approx3\,{\mu\rm G}$. As shown in Fig. \ref{fig:sed}, the $\gamma$-ray emission remains detectable by current and upcoming VHE and UHE facilities, for a source at 3 kpc, long after the MQ shut down. 

Although we have focused on systems with powerful, well–collimated jets reminiscent of FR~II radio galaxies (e.g., S26 in NGC~7793; \cite{Abaroa_etal2024(S26)}), a similar reasoning applies to MQs with slower jets and prominent recollimation shocks, such as SS~433. In those sources, particle acceleration is likely to occur at the recollimation region rather than at terminal hotspots, but the resulting cocoon and CR reservoir would behave in a similar way \citep{SS433-2024Sci}. The parameter space of MQRs therefore encompasses both FR~II–like and FR~I–like MQs.

In contrast to supernova remnants (SNRs), which typically exhibit well-defined shell-like morphologies and multiwavelength counterparts (radio and often X-rays), MQRs are expected to appear as more diffuse, centrally underluminous bubbles with no prominent outer shock once the jet activity has ceased. The $\gamma$-ray emission would trace dense gas irradiated by relic CRs, potentially leading to spatial offsets between the $\gamma$-ray source and the geometric center of the MQR. 
Moreover, energy-dependent diffusion predicts a systematic spectral change with increasing distance from the reservoir, a feature that could help distinguish MQRs from young SNR PeVatrons, where the highest-energy particles are confined near active shock fronts.

A rough estimate of the expected Galactic population of MQRs can be obtained from the known number of high-mass X-ray binaries hosting super-Eddington phases. If the active MQ phase lasts $5\times 10^4\,$yr and the cocoon can retain PeV particles for $\gtrsim 10^5\,$yr, the fossil stage may outlive the active phase by a factor of a few. Given that several tens of high-mass MQs are currently known in the Galaxy, and accounting for past episodes over Galactic timescales, the number of MQRs could plausibly reach several tens to at most a few hundred objects, depending on the duty cycle of super-Eddington accretion and binary survival rates. This suggests that MQRs may represent a non-negligible fraction of the unidentified Galactic PeVatrons detected by LHAASO.

In summary, we propose that remnants of extinct super–Eddington MQs may constitute an overlooked population of Galactic PeVatrons. Their cocoons act as extended CR reservoirs capable of illuminating nearby dense gas over long timescales, generating VHE and UHE $\gamma$–ray emission with no obvious compact counterpart. Given that several active super–Eddington MQs are known in the Galaxy, the existence of their fossil cousins appears unavoidable. A systematic search for faint, large–scale radio structures around clusters of unidentified LHAASO sources, combined with detailed modeling of their $\gamma$–ray spectra and morphology, may provide the key to unveiling this hidden population of ghost MQs.

\acknowledgments{LA thanks the Universidad Nacional de La Plata and the ICTP/SAIFR. GER and VBR were funded by PID2022-136828NB-C41/AEI/10.13039/501100011033/ and through the ``Unit of Excellence María de Maeztu'' award to the Institute of Cosmos Sciences (CEX2019-000918-M, CEX2024-001451-M). VB-R is Correspondent Researcher of CONICET, Argentina, at the IAR.}

\bibliographystyle{jhep}
\bibliography{main}


\end{document}